\documentclass[12pt]{article}
\usepackage{amssymb}
\usepackage{graphicx}
\usepackage{amsmath}

\def\e{{\rm e}}

\def\d{\partial}
\def\l{\left(}
\def\r{\right)}

\newcommand{\be}{\begin{equation}}
\newcommand{\ee}{\end{equation}}
\newcommand{\bea}{\begin{eqnarray}}
\newcommand{\eea}{\end{eqnarray}}
\newcommand{\bg}{\begin{gather}}
\newcommand{\eg}{\end{gather}}
\newcommand{\bseq}{\begin{subequations}}
\newcommand{\eseq}{\end{subequations}}

\newcommand{\arctg}{\mathop{\rm arctg}\nolimits}

\renewcommand{\ln}{\mathop{\rm ln}\nolimits}

\begin{document}
\begin{flushright}
CERN-PH-TH/2008-073
\end{flushright}

\begin{center}
{\Large \bf Self-accelerated brane Universe 
with warped\\ extra dimension}

\medskip
D.S.~Gorbunov$^{a}$\footnote{gorby@ms2.inr.ac.ru},
S.M.~Sibiryakov$^{b,a}$\footnote{Sergey.Sibiryakov@cern.ch, 
~sibir@ms2.inr.ac.ru} 
\\
\medskip
$^a${\small
Institute for Nuclear Research of the Russian Academy of Sciences,\\  
60th October Anniversary prospect, 7a, 117312 Moscow, Russia.}\\
$^b${\small Theory Group, Physics Department, CERN, CH-1211 Geneva 23,
  Switzerland.}
\end{center}
\vspace{0.5cm}

\begin{abstract}

We propose a cosmological model which exhibits the phenomenon of
self-ac\-ce\-le\-ra\-tion: the Universe is attracted to the phase of
accelerated expansion at late times even in the absence of the
cosmological constant. The self-acceleration is inevitable in the
sense that it cannot be neutralized by any negative 
explicit cosmological
constant. The model is formulated in the framework of brane-world
theories with a warped extra dimension. 
The key ingredient of the model is the brane-bulk energy transfer
which is carried by bulk vector fields with a 
$\sigma$-model-like
boundary condition on the brane. 
We explicitly find the 5-dimensional metric corresponding to the
late-time de Sitter expansion on the brane;
this metric describes an
AdS$_5$ black hole with growing mass.  
The present value of the
Hubble parameter implies the scale of new physics of order 1~TeV,
where the proposed model has to be replaced by putative UV-completion.  
The mechanism leading to the self-acceleration has AdS/CFT
interpretation as occurring due to specific dynamics of conformal matter
interacting with external ``electric'' fields. The 
Universe expansion history predicted by the model 
is distinguishable from the standard
$\Lambda$CDM cosmology.

\end{abstract}

%%%%%%%%%%%%%%%%%%%%%%%%%%%%%%%%%%%%%%%%%%%%%%%%%%%%%%%%%%%%%%%%%%%%%%%%
\section{Introduction and summary}

One of the open questions of the contemporary physics is the nature of 
 dark energy.  The latter term is conventionally used for the
unknown substance which is believed to be responsible for the observed
acceleration of the cosmological expansion. In the minimal scenario, the
dark energy is identified with the cosmological constant (CC).  
However, the energy scale associated with the value of the CC required
to explain the observed dark energy density~\cite{Yao:2006px}, 
\be
\label{CCobs}
\rho_{DE}=(0.43\cdot 10^{-10}~\mathrm{GeV})^4\;,
\ee
is much smaller than
any other fundamental energy scale existing in particle
physics. Thus, one faces the problem of explaining this strong
hierarchy \cite{Weinberg:1988cp}. All particle condensates
(i.e., those associated with spontaneous symmetry breaking)
contribute to CC 
but have to almost cancel out. Moreover, the 
CC receives contributions from quantum
fluctuations of all fields so that a natural value for it would be
\be
\label{CCnat}
\rho^{(\mathrm{natural})}_\Lambda\sim \Lambda_{UV}^4\;,
\ee  
where $\Lambda_{UV}$ is the ultra-violet cutoff of the theory. The
validity of the Standard Model of particle physics up to energies of
order 1~TeV implies that $\Lambda_{UV}\gtrsim 1$~TeV. The latter value
of $\Lambda_{UV}$ when substituted into (\ref{CCnat}) gives the
natural value of the CC much larger than the one consistent with
observations. 

Of course, Eq.~(\ref{CCnat}) is not rigorous: quantum corrections to
CC are divergent and can be formally put to any value by
renormalization.  There exist several proposals to cancel the CC (for
reviews see, e.g.,
Refs.~\cite{Weinberg:1988cp,Nobbenhuis:2004wn,Polchinski:2006gy}).
They can be split in two groups. One group involves proposals
based on relaxation mechanisms
\cite{Abbott:1984qf,Brown:1988kg,Rubakov:1999aq,Dolgov:2002yw,Steinhardt:2006bf}.
These proposals suffer from the ``empty Universe'' problem
(see, e.g., \cite{Polchinski:2006gy}): the relaxation mechanisms drive
the Universe to the state practically devoid of matter which is
inconsistent with the Hot Big Bang. It is not clear for the moment
whether this problem can be consistently resolved without invoking
anthropic selection \cite{Weinberg:1987dv}. Another group involves
symmetry \cite{Linde:1989dr} or quantum gravity
\cite{Hawking:1984hk,Coleman:1988tj} arguments. However, these
proposals adjust the CC somewhat too well: they predict that the CC is
exactly zero. The problem is now reversed: obtaining the Universe
undergoing the late-time de Sitter expansion requires fine-tuning.

In this context one would like to construct a setup which excludes the
Minkowski space-time from the possible asymptotics of the history of
the Universe {\it irrespectively} of the value of the CC. Namely, we
envisage the following picture. The Universe is driven to the de
Sitter phase at late times not by the CC but by some dynamical
mechanism.  The Hubble parameter of the asymptotic de Sitter
space-time may depend on the value of the CC but is always bounded
from below by a tiny positive value. This reduces the CC problem to
its old form: the need to explain the cancellation of the CC
(admittedly, this is the most difficult task) without the risk of
being over-precise.  An additional requirement one would like to
impose on the dynamical mechanism leading to the late-time
acceleration of the Universe is that it should not invoke the small
energy scale entering Eq.~(\ref{CCobs}) as an input and should not
require any particular initial conditions for the evolution of the
Universe. It is natural to refer to the scenario described above as
``self-accelerated cosmology''.

The model that came closest to the realization of the self-accelerated
scenario is the DGP model \cite{Dvali:2000hr}. It is formulated in 
flat (4+1)-dimensional space-time with our Universe being
represented by a 3-dimensional brane; the key ingredient of the model
is the induced gravity term on the brane. Still, the self-accelerated
branch of this model suffers from the presence of ghost fields
\cite{Luty:2003vm,Gorbunov:2005zk}, i.e. excitations with negative
kinetic term which signal inconsistency of the theory.

In this paper we propose another model which realizes the idea of
self-accelerated cosmology. The model is formulated in the context of
brane-world scenario with warped extra dimensions
\cite{Randall:1999vf}.  We do not consider the induced gravity term
on the brane. Instead, we introduce an energy exchange between the
brane and the bulk carried by a triplet of $U(1)$ bulk vector fields
\cite{Gorbunov:2005dd}.  Then, the self-acceleration mechanism can be
summarized as follows. As the brane expands, it gives rise to an
energy flow from the brane into the bulk. This energy flow, in its
turn, results in the formation of a bulk black hole with growing
mass. The latter back-reacts on the brane and maintains the brane
Hubble parameter constant at late times.\footnote{
The increase of the brane Hubble parameter by a bulk black hole is a
known effect:
a static black hole acts as ``dark
radiation'' term in the Friedman equation \cite{Binetruy:1999hy}. 
For a constant mass black
hole this effect dies out at late times while in our case it persists
due to the growth of the black hole mass.    
} 
Similar ideas
were expressed previously in Ref.~\cite{Apostolopoulos:2005at}; our
model provides explicit realization of these ideas in the framework of
field theory.
The brane-bulk energy exchange
is kept at the level sufficient to provide the self-accelerated
cosmology by a specific Dirichlet boundary condition for the vector
fields which forces their squared values on the brane to be equal to a
fixed parameter $M^2$, see Eq.~(\ref{ABC}). Note that existence of a
continuous energy flux from the brane implies violation of the null
energy condition by the brane energy-momentum tensor. This
property is dangerous as it can lead to instabilities
\cite{Dubovsky:2005xd}. Below we briefly discuss this problem and a
possible way to resolve it. A more detailed treatment of this issue
will be presented elsewhere \cite{Perturbations}. In the present paper
we concentrate on the cosmological applications of our model.

When the explicit CC on the brane
vanishes, 
the self-accelerated expansion regime at late times
corresponds to the value of the would-be CC given by the seesaw
formula, 
\be
\label{seesaw}
\rho_{\Lambda_{eff}}\sim \l M^2/M_{Pl}\r^4\;. 
\ee
The present dark energy density
(\ref{CCobs}) corresponds to
$M$ of order 1 TeV. 
It is important to stress that self-acceleration
mechanism we propose is not
equivalent to an implicit introduction of a small positive CC. We
check this 
in two ways. First, we find that in the absence of an explicit CC and
matter on the brane the system admits, apart from the self-accelerated
solution, also a static solution with Minkowski metric on the
brane. However, in contrast to the self-accelerated solution, the
Minkowski solution is unstable and is not an attractor of the
cosmological evolution.  This solves the problem of naturalness of the
initial conditions leading to late-time acceleration: in our model the
Universe always approaches the de Sitter space at late times.  Second,
we show that the self-acceleration mechanism cannot be compensated by
an explicit introduction of a negative CC on the brane. Namely,
depending on the value of the negative CC, the Universe either
undergoes the accelerated expansion with essentially the same Hubble
parameter corresponding to (\ref{seesaw}), or collapses. The latter
behavior, the collapse for large and negative explicit CC,  is
similar to the collapse of the Universe in the case of negative CC
in the standard Friedman cosmology. Presumably, a (yet unknown)
mechanism which cancels the large positive CC, excludes the
collapsing cosmological solutions as well,
cf. \cite{Hawking:1984hk,Coleman:1988tj}.

The self-acceleration mechanism realized in our model has a
transparent interpretation in term of the AdS/CFT correspondence
\cite{Aharony:1999ti}.  In this language the acceleration is driven by
the conformal matter. The cooling of this matter due to the expansion
of the Universe is compensated by an energy inflow due to work done by
external ``electric'' fields on the CFT. The latter do not decay
rapidly in the expanding Universe in our model. Technically, this is
achieved by the specific $\sigma$-model-like conditions on the vector
fields, Eq.~(\ref{ABC}).  The
CFT interpretation of the self-acceleration mechanism suggests
phenomenological generalizations of the model. The cosmological
expansion equation in this class of models is distinct
from the standard $\Lambda$CDM case, thus providing falsifiable
predictions. A careful analysis of this and other phenomenological
consequences of our model and its generalizations is left for future
work.

The paper is organized as follows. 
In Sec.~\ref{Sec:2} we introduce the 
setup and identify its features that are relevant for cosmology. 
Section~\ref{Sec:3} contains qualitative analysis of the cosmological
evolution in the model.  
A more rigorous analysis is
performed in Sec.~\ref{Sec:4} where an explicit solution describing
a self-accelerated universe is given. In Sec.~\ref{Sec:5} we
describe the interpretation of this cosmological solution in the spirit
of the AdS/CFT correspondence. 
In Sec.~\ref{Sec:7} we discuss open issues related to our work.
Appendix \ref{App:A00} presents an explicit example of 
a brane Lagrangian giving rise to the self-accelerated cosmology. 
Appendices \ref{App:A0} and \ref{App:A} contain 
technical details.

%%%%%%%%%%%%%%%%%%%%%%%%%%%%%%%%%%%%%%%%%%%%%%%%%%%%%%%%%%%%%%%%%%%%%%%%%%

\section{Setup}
\label{Sec:2}
We consider a brane-world model based on the setup analogous to that of 
Ref.~\cite{Randall:1999vf}. All the fields
describing the observable matter 
are supposed to be localized on a 3-brane. 
%whose worldvolume serves as
%a boundary of 5-dimensional anti-de Sitter (AdS) 
%bulk. 
The action of the model is the sum
of bulk and brane parts,
\[
S=S_{bulk}+S_{brane}\;.
\]
The bulk sector consists of gravity described by the standard
Einstein--Hilbert action with negative cosmological constant $\Lambda_5$ 
and of a triplet of Abelian bulk vector fields 
$A_M^a$, $a=1,2,3$. Thus, for the bulk action we 
have,\footnote{Our convention for the metric signature is 
  $(+,-,-,\ldots)$; we use upper-case Latin letters $M,N,\ldots$ 
  for 5-dimensional indices, the Greek letters $\mu,\nu,\ldots$ for
  4-dimensional indices in the tangent space to the brane, and 
  lower-case Latin letters $i,j,\ldots$ for the 3-dimensional spatial
  indices.}  
\[
S_{bulk}=\int d^5x\sqrt{g}
\left(-\frac{R}{16\pi G_5}-\Lambda_5
-\frac{1}{4e^2}F_{MN}^aF^{a\,MN}\right)\;,
\] 
where
\[
F_{MN}^a=\d_MA_N^a-\d_NA_M^a\;.
\]
The brane action is taken in the following form,
\be
\label{Sbrane}
S_{brane}=\int d^4x\sqrt{-\bar g}\,\big(-\sigma+{\cal L}_{SM}+
{\cal L}_{HS}[A_\mu^a,\ldots]\big)\;.
\ee 
Following \cite{Randall:1999vf} we impose $Z_2$ reflection symmetry
across the brane.
In the above brane action $\sigma$ is the tension of the brane, 
${\cal L}_{SM}$ represents 
the Standard Model Lagrangian (possibly, including dark matter
sector), and ${\cal L}_{HS}$ is a boundary Lagrangian 
for the vector fields. 
The dots 
represent possible dependence of
${\cal L}_{HS}$ on some
additional fields localized on the brane. 
We will refer to the sector described by ${\cal L}_{HS}$ as
the ``hidden sector'': we assume that
there are no direct interactions of the vector fields $A^a_\mu$ as
well as of the other fields entering into  ${\cal L}_{HS}$ 
(except the metric
$g_{\mu\nu}$) with the Standard
Model fields. 

For  the brane tension we write
\[
\sigma=\sqrt{-3\Lambda_5/4\pi G_5}+\rho_\Lambda\;.
\]
As we will see shortly $\rho_\Lambda$ plays the role of
the 4-dimensional CC. We assume that there is some mechanism 
ensuring that $\rho_\Lambda$ is much smaller than its natural value
(\ref{CCnat}). Without knowledge of the details of this
mechanism we cannot be more concrete and say what value 
of $\rho_\Lambda$ it predicts.
A reasonable possibility is
that $\rho_\Lambda$ is set equal to
zero but other options cannot be excluded.
So
we will keep $\rho_\Lambda$ explicitly in the formulas in order to be
able to control its
effect on the cosmological evolution.
 
The precise form of the hidden
sector Lagrangian is not important for our purposes. 
For the study of cosmology it is sufficient to require the following
properties: 

(1) Variation of ${\cal L}_{HS}$ gives
the following boundary condition for the vector fields on the brane:
\be
\label{ABC}
\bar g^{\mu\nu}A^a_\mu A^b_\nu\Big|_{brane}=-M^2\delta^{ab}\;, 
\ee
where $\bar g^{\mu\nu}$ is the induced metric on the brane, and $M$ is
some parameter of dimension of mass.  Though the condition (\ref{ABC})
has a generally covariant form
it can be satisfied only if
the vector fields are non-zero so that the local Lorentz symmetry is
spontaneously broken.
Note also that Eq.~(\ref{ABC}) explicitly
breaks the $U(1)$ gauge symmetries of the bulk action.  Having in mind
an analogy with $\sigma$-models of particle physics one expects that 
the use of the condition (\ref{ABC}) is justified at energies below
$M$ which are relevant for late-time cosmology. On the other hand,
at energies above $M$ some UV-completion of the theory is 
required.\footnote{In fact, the scale $M$ is an {\it upper bound } on
  the UV-cutoff of 
the theory. The actual value of the cutoff can be lower in a given
model of the brane Lagrangian. This issue goes beyond the scope of 
the present article.}

(2) The energy density of the hidden sector matter dilutes,
$\rho_{HS}\to 0$, as the brane undergoes the cosmological
expansion. In other words, we assume that the hidden sector as well as
the Standard Model sector does not
contain any CC. 
In
order not to contradict the observational data $\rho_{HS}$ should
dilute at the same rate or faster than the energy density of dust.

An explicit example of the
Lagrangian ${\cal L}_{HS}$ satisfying the requirements (1), (2)
is presented in appendix \ref{App:A00}. 

We make the following assumptions about the
order-of-magnitude values of the parameters of our setup. 
The bulk physics is governed by the energy scale of order or
slightly below the Planck mass. Thus, we have
\be
\label{assump}
G_5\sim M_{Pl}^{-3}~,~~~e^2\sim M_{Pl}^{-1}~, 
~~~\Lambda_5\sim M_{Pl}^5\;.
\ee
On the other hand, the
scale $M$ entering into Eq.~(\ref{ABC}) is assumed to be much smaller
than 
the Planck mass. Below we will take $M\sim \text{1 TeV}$ in order to
fit the observed acceleration rate of the Universe.

%%%%%%%%%%%%%%%%%%%%%%%%%%%%%%%%%%%%%%%%%%%%%%%%%%%%%%%%%%%%%%%%%%%%

\section{Cosmology: qualitative discussion}
\label{Sec:3}

\subsection{Cosmological Ansatz}
\label{Sec:3.1}

\begin{figure}[!htb]
\begin{center}
%\begin{picture}(0,0)%
\includegraphics[width=0.5\textwidth]{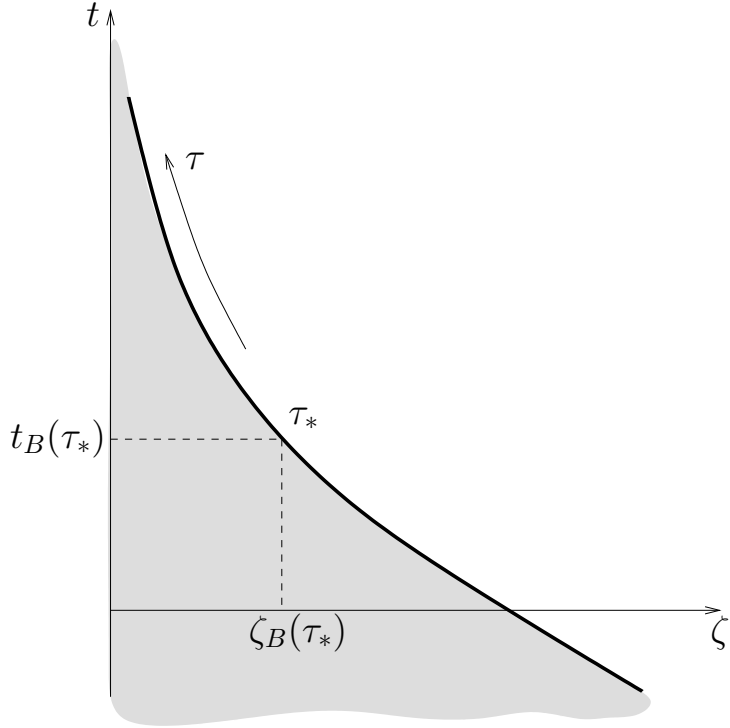}%
%\end{picture}%
\setlength{\unitlength}{3947sp}%
\begingroup\makeatletter\ifx\SetFigFont\undefined%
\gdef\SetFigFont#1#2#3#4#5{%
  \reset@font\fontsize{#1}{#2pt}%
  \fontfamily{#3}\fontseries{#4}\fontshape{#5}%
  \selectfont}%
\fi\endgroup%
%\begin{picture}(5724,6863)(-311,-5787)
\begin{picture}(0,0)(0,0)
{\large
\put(-4020,4400){$t$}
\put(-3400,3500){$\tau$}
\put(-2750,1900){$\tau_*$}
\put(-3000,530){$\zeta_B(\tau_*)$}
\put(-4500,1750){$t_B(\tau_*)$}
\put(-100,530){$\zeta$}
}
\end{picture}%
\caption{The coordinate frame used in the text. Thick curve
  shows the trajectory of the brane. The space-time considered in the
  text consists of two copies of white region (above the 
thick curve) glued across the brane trajectory
  and related by $Z_2$ symmetry. Gray region is cut off.
\label{brane-embedded}
}
\end{center}
\end{figure}

We now turn to the discussion of cosmology in the model described
above. The general Ansatz for the metric and vector fields, 
invariant under
3-dimensional translation and rotations, has the form 
\begin{align}
\label{start1}
&ds^2=F(t,\zeta)(dt^2-d\zeta^2)-r^2(t,\zeta){(dx^i)}^2;\\
\label{start2}
&A^a_i=M A(t,\zeta)\,\delta^a_i~,~~~~~A^a_t=A^a_\zeta=0.
\end{align}
Equations~(\ref{start2}) imply that the vector fields $A^a_M$ form an
orthogonal spacelike triad; such a configuration is invariant under
3-dimensional rotations supplemented by simultaneous rotations in the
internal space. 
Embedding of the brane into the space with 
metric (\ref{start1}) is described by
a set of functions $x^M(y^\mu)$. Choosing $x^i=y^i$ one is
left with two functions, $t_B(\tau)$ and
$\zeta_B(\tau)$, describing the trajectory of the brane on 
the $(t,\zeta)$ plane. 
Here $\tau$ is a time parameter on the brane, see
Fig.~\ref{brane-embedded}.  
It is convenient to choose $\tau$ to be the cosmological time. Then 
the induced metric on the brane takes the Robertson--Walker form
\be
\label{brmetric}
d\bar s^2=d\tau^2-a^2(\tau) (dx^i)^2\;.
\ee
Here $a(\tau)=r(t_B(\tau),\zeta_B(\tau))$, and the functions $t_B$,
$\zeta_B$ obey the condition
\[
F(t_B,\zeta_B)\big(\dot t_B^2-\dot\zeta_B^2\big)=1\;,
\]
where dot denotes derivative with respect to $\tau$. The motion of
the brane is affected by the bulk metric via the junction
conditions. Imposing $Z_2$ reflection symmetry
across the brane one obtains,
\bseq
\label{junc}
\begin{align}
\label{junc1}
&4\pi G_5 T_{(br)\,\tau}^\tau=-3\left(\frac{\d_tr}{r}\dot\zeta_B
+\frac{\d_\zeta r}{r}\dot t_B\right)\;,\\
\label{junc2}
&4\pi G_5
T_{(br)\,i}^j=-\left[\left(\frac{2\d_tr}{r}+\frac{\d_tF}{2F}\right)
\dot \zeta_B
+\left(\frac{2\d_\zeta r}{r}+\frac{\d_\zeta F}{2F}\right)\dot t_B
+\frac{1}{\sqrt F\dot t_B}\frac{d}{d\tau}(\sqrt F\dot\zeta_B)\right]
\delta_i^j\;.
\end{align}
\eseq
Finally, the boundary
conditions (\ref{ABC}) relate the boundary value of the function 
$A(t,\zeta)$ to the scale factor of the brane metric,
\be
\label{Abrane}
A(t_B(\tau),\zeta_B(\tau))=a(\tau)\;.
\ee 
This equation implies that the cosmological expansion of the
brane acts as a source for the bulk vector fields. As we
will see, this lies at the origin of the self-acceleration mechanism
in our model.

The junction conditions (\ref{junc}) are simplified if
the Hubble parameter $H\equiv \dot a/a$ on the brane is
small compared to the curvature of the bulk. In this case one expects 
the 
distortion of the bulk due to the vector fields emitted by the brane to be
small near the brane. Then the metric in the vicinity of the brane
is approximately that of 
AdS$_5$:
\be
\label{slow}
F=\frac{1+2\phi(t,\zeta)}{(k\zeta)^2}~,~~~~
r=\frac{1+\psi(t,\zeta)}{k\zeta}\;,
\ee 
where $k\equiv\sqrt{-4\pi G_5\Lambda_5/3}$ is the inverse AdS radius, and 
$\phi, \psi\ll 1$. Expanding Eqs.~(\ref{junc}) to the
linear order in $\phi,\psi$ and to quadratic order in $H/k$ one finds
the equations governing the cosmological expansion of the brane,
\begin{align}
\label{newfried}
&H^2=\frac{8\pi G_N}{3}(\rho_\Lambda+\rho)+2k^2(\phi+\zeta\d_\zeta\psi)\big|_{brane}\;,\\
\label{accel}
&\dot H=-4\pi G_N(\rho+p)+k^2(\zeta\d_\zeta\phi-
\zeta\d_\zeta\psi)\big|_{brane}\;.
\end{align}
Here $G_N=G_5k$ is the four-dimensional Newton constant, and $\rho$,
$p$ are the energy density and pressure of the brane matter 
(including both the Standard Model and hidden sector fields).  
Without the terms on the r.h.s. containing the functions 
$\phi$, $\psi$, equations (\ref{newfried}), (\ref{accel}) would
describe the standard cosmology with the cosmological constant
$\rho_\Lambda$. The above terms are responsible for the modification
of the brane cosmology in our model; they
are to be found from the bulk dynamics.

Let us comment on the validity of the approximation 
(\ref{slow}). We will check below that it is indeed
valid in the vicinity of the brane if
\[
H/k\ll 1\;.  
\] 
The latter inequality is readily satisfied in the
regime of interest: eventually we want $H$ to be comparable to the
present Hubble parameter of the Universe, while $k$ is assumed to be
of order $M_{Pl}$.  On the other hand the form (\ref{slow}) is
inapplicable far from the brane: we will see shortly that the effect
of the vector fields on the space-time geometry increases deep into the
bulk and leads to strong deviation of the metric from that of AdS$_5$.

\subsection{Vector fields in external metric}
Let us analyze
the behavior of the vector fields in the curved metric
(\ref{start1}). Using the cosmological Ansatz (\ref{start2})
one derives their equation of motion
\be
\label{vector}
\d_t(r\d_tA)-\d_\zeta(r\d_\zeta A)=0
\ee
and the expressions for the components of their
energy--momentum tensor:
\bseq
\label{Tvect}
\begin{align}
\label{Tvect1}
&T_{tt}=T_{\zeta\zeta}=
\frac{3M^2}{2e^2r^2}\big((\d_tA)^2+(\d_\zeta A)^2\big)\;,\\
\label{Tvect2}
&T_{t\zeta}=\frac{3M^2}{e^2r^2}\d_tA\d_\zeta A\;,\\
\label{Tvect3}
&T_{ij}=\delta_{ij}\frac{M^2}{2e^2F}
\big((\d_t A)^2-(\d_\zeta A)^2\big)\;.
\end{align}
\eseq
Note that non-vanishing value of $T_{t\zeta}$ implies non-vanishing
energy flux along the $\zeta$-direction carried by the
vector fields. 

It is instructive to neglect for the moment the backreation of the
vector fields on the metric and consider their evolution in the
external AdS$_5$ geometry. Accordingly, one sets
\be
\label{AdSmetr}
F=\frac{1}{(k\zeta)^2}~,~~~r=\frac{1}{k\zeta}\;.
\ee
Let us also forget about the brane for a while and consider the
Dirichlet boundary condition for the vector field 
{\em at the AdS boundary}, 
\be
\label{dirichlet}
A(t,\zeta)\to A^{(0)}(t)~,~~~\zeta\to 0^+\;. 
\ee
The relevance of this problem for the original cosmological setup will
become clear shortly. Performing the Fourier decomposition,
\[
A(t,\zeta)=\int\frac{d\omega}{2\pi}
\tilde A(\omega,\zeta)\e^{-i\omega t}\;,
\] 
and solving Eq.~(\ref{vector}) one finds
\be
\label{solution}
\tilde A(\omega,\zeta)=
\frac{i\pi}{2}\omega\zeta H_1^{(1)}(\omega\zeta)\tilde A^{(0)}(\omega)\;.
\ee 
This expression suggests two important observations. 

First, the
vector fields are almost constant 
at distances $\zeta\ll 1/\omega$ from the
AdS boundary,
\[
\frac{\tilde A(\omega,\zeta)-\tilde A(\omega,0)}
{\tilde A(\omega,0)}\ll 1~,~~~\text{for}~~\omega\zeta\ll 1\;.
\]
The latter condition is satisfied by the position $\zeta_B$ of the
brane in our cosmological setup. Indeed, the characteristic
frequency entering the problem can be estimated as 
\be
\omega\sim \frac{1}{A(t_B,\zeta_B)}\frac{dA(t_B,\zeta_B)}{dt_B}\;.
\ee
Using the relation (\ref{Abrane}) between the boundary value of the vector
field and the scale factor on the brane one obtains,
\be
\omega\zeta_B\sim\frac{\zeta_B}{a}\frac{da}{dt_B}=
\zeta_B\frac{da}{d\tau}=\frac{H}{k}\ll 1\;,
\ee
where in the last equality we substituted $a=1/(k\zeta_B)$. Therefore one
can simplify the set of equations determining the cosmological solution
by
replacing Eq.~(\ref{Abrane}) with
\be
\label{Abrane1}
A(t_B(\tau),0)=a(\tau)\;.
\ee
Instead of being formulated on the moving brane, 
the boundary condition for the vector
field is now imposed on the fixed timelike surface --- the AdS boundary.

Second, the formula (\ref{solution}) can be used to estimate the
backreaction of the vector fields on the metric. For example, let us
consider the trace of energy--momentum tensor of these fields. From
Eqs.~(\ref{Tvect}) one obtains that the contribution of a given
harmonic at $\omega\zeta\gg 1$ has the form
\[
T_M^M=-\frac{3\pi M^2k\omega 
|\tilde A^{(0)}(\omega)|^2}{64e^2}(k\zeta)^3\;.
\]
We see that $T_M^M$ grows into the bulk. Thus, at large 
distance from the AdS boundary the backreaction of the vector
fields on the space--time geometry gets large and should be properly
taken into account. We proceed to this task.

\subsection{Account for backreaction}
\label{Sec.3.3}
The above analysis suggests the following physical picture.
The cosmological expansion of the brane generates the bulk vector
fields.
The $T_{t\zeta}$ component of the energy--momentum tensor of these
fields is non-zero, so they
carry a flux of energy out of the brane. Due to the
warping of the bulk metric 
the energy density of the flux grows into the
bulk leading to formation of a black hole\footnote{Another possibility could
  be formation of a naked singularity. According to the ``cosmic
  censorship'' hypothesis we assume that this is not the case. This 
assumption will be confirmed 
  by explicit solution of the Einstein equations in Sec.~\ref{Sec:4}.}.
It is known \cite{Binetruy:1999hy} 
that the presence of a constant mass black hole in the
bulk affects the cosmology on the brane via the so called ``dark
radiation'' term which dilutes as $1/a^4$ at late times. In our case 
the mass of the black hole is not constant: it grows with
time due to the continuous inflow of energy from the brane. This leads
to nontrivial consequences at late times.

To determine the growth rate of the black hole mass rigorously, one
needs to solve the system consisting of equations of motion
for the vector fields and the bulk Einstein equations. This
will be done in the next section for the case of the self-accelerated
cosmological Ansatz.  For the moment we adopt a simplified strategy
which allows to catch the qualitative features of the resulting
cosmology. Namely, we will solve for the vector fields in the
background of a {\em static} black hole. The energy flux given by this
solution will be then inserted into the energy conservation equation
in order to calculate the dependence of the black hole mass on
time. Finally, we will use the metric of the black hole with time
dependent mass to evaluate the bulk terms in the effective Friedman
equation (\ref{newfried}) on the brane.

A bulk black hole\footnote{Strictly speaking, the configuration we are
considering should be called black brane as it is extended along 3
spatial direction. However, we prefer the more conventional term
``black hole''.}  with mass ${\cal M}$ is described by the
AdS--Schwarzschild metric, \be
\label{adsschw}
ds^2=F(r)dt^2-\frac{dr^2}{k^2F(r)}-r^2dx_i^2\;,
\ee
where
\be
\label{Fr}
F(r)=r^2-\frac{r_h^4}{r^2}\;,
\ee
and the Schwarzschild radius $r_h$ is related to the black hole mass 
${\cal M}$ as follows, 
\be
\label{rh}
r_h=\left(\frac{8 G_5k^2{\cal M}}{3\pi}\right)^{1/4}\;.
\ee
This metric can be cast into the form (\ref{start1}) with 
$\zeta$ determined by the differential equation
\[
\frac{d\zeta}{dr}=-\frac{1}{kF}\;.
\]
Explicitly,
\be
\label{rzeta}
\zeta=-\frac{1}{4kr_h}
\ln\left[\frac{r-r_h}{r+r_h}\right]-
\frac{1}{2kr_h}\left[\arctg{\frac{r}{r_h}}-\frac{\pi}{2}\right]\;.
\ee
In this background the vector equation (\ref{vector}) takes
the form
\be
\label{vecschw}
-\d_\zeta^2A-\frac{d\ln r}{d\zeta}\d_\zeta A+\d_t^2A=0\;.
\ee
To understand the structure of the solutions to this equation 
let us solve it with the Dirichlet condition on the AdS boundary,
\be
\label{dirichlet1}
A(t,\zeta=0)=A_\omega^{(0)}\e^{-i\omega t}\;.
\ee
One looks for the solution in the form $A=A_\omega(\zeta)\e^{-i\omega t}$. 
Let us consider the case of small frequencies, 
\be
\label{smallfr}
\omega\ll kr_h\;.
\ee
Then, the
solution can be constructed in the following way: one solves
Eq.~(\ref{vecschw}) in two regions, $\zeta\ll 1/\omega$ and $\zeta\gg
1/kr_h$, and matches the solutions in the intersection of these regions,
$1/kr_h\ll\zeta\ll 1/\omega$. 
At small values of $\zeta$ the function
$A_\omega(\zeta)$ varies slowly with $\zeta$ and one can
approximate its time derivatives by their values on the AdS boundary,
\be
\label{approx}
\d_t^2A_\omega\approx -\omega^2A_\omega^{(0)}\;.
\ee
Equation (\ref{vecschw}) yields  
\be
\label{dzA}
\d_\zeta A_\omega=
-\frac{\omega^2A^{(0)}_{\omega}}{4kr}
\left(-\ln\left[\frac{r^4}{r^4_h}-1\right]+C_\omega\right)\;,
\ee
where $C_\omega$ is the integration constant. The range of validity of this
formula is restricted by the condition 
(\ref{approx}) which translates into
\be
\label{smaldz}
\zeta\ll\frac{1}{\omega}~,~\frac{1}{\omega \sqrt{C_\omega}}\;.
\ee
In the region $\zeta\gg 1/kr_h$ the expression (\ref{dzA}) takes the
form,
\be
\label{A0inter}
\d_\zeta A_\omega=-\frac{\omega^2 A_\omega^{(0)}}{4kr_h}
\left(-3\ln 2-\frac{\pi}{2}+C_\omega\right)-\omega^2A_\omega^{(0)}\zeta\;.
\ee
On the other hand, 
at $\zeta\gg 1/kr_h$ the radial function $r(\zeta)$ gets stabilized at the
value of the Schwarz\-schild radius, and the second term in
Eq.~(\ref{vecschw}) vanishes. So, in this region the solutions of 
Eq.~(\ref{vecschw})
are free waves. We take the
outgoing solution
\[
A_\omega(\zeta)=A_\omega^{(1)}\e^{i\omega\zeta}\;.
\] 
At $\zeta\ll 1/\omega$ this formula gives
\[
\d_\zeta A_\omega=i\omega A^{(1)}_\omega-\omega^2 A^{(1)}_\omega\zeta\;.
\]
Comparing this expression with Eq.~(\ref{A0inter}), one obtains
\bseq
\label{match}
\begin{align}
\label{match1}
&A_\omega^{(0)}=A_\omega^{(1)}\;,\\
\label{match2}
&C_\omega=-i\frac{4kr_h}{\omega}+3\ln 2+\frac{\pi}{2}
\approx -i\frac{4kr_h}{\omega}\;.
\end{align}
\eseq
Note, that though $C_\omega$ is large, $|C_\omega|\gg 1$, the second
inequality in (\ref{smaldz}) is still compatible with $\zeta\gg
1/kr_h$. Substitution of (\ref{match2}) into Eq.~(\ref{dzA}) yields
at $\zeta\ll 1/\omega$,
\be
\label{dza2}
\d_\zeta A_\omega= \frac{\omega^2A^{(0)}_{\omega}}{4kr}
\ln\left[\frac{r^4}{r^4_h}-1\right]+\frac{i\omega
r_h}{r}A^{(0)}_{\omega}\;.  
\ee 
Due to the condition (\ref{smallfr})
the first term on the r.h.s. of Eq.~\eqref{dza2} is small in
comparison with the second one if the logarithm is not too large. We
will see below that this condition is satisfied by 
the solution we are
interested in.

It is straightforward to generalize the formula (\ref{dza2}) 
to the case of an
arbitrary dependence of the Dirichlet boundary condition on time,
\[
A(t,\zeta=0)=A^{(0)}(t)\;.
\]
Omitting the first term on the r.h.s in Eq.~(\ref{dza2}) one obtains
\be
\label{dza3}
\d_\zeta A=-\frac{r_h}{r}\d_t A^{(0)}\;.
\ee
This expression is valid when the characteristic frequency 
$\omega\sim \d_tA^{(0)}/A^{(0)}$ satisfies the inequality (\ref{smallfr}).

The next step is to determine the change of the metric due to
the inflow of energy from the brane into the black hole. The increase
of the black hole mass is given by the formula  
\[
\d_t{\cal M}=\frac{2\pi^2}{k^3}\sqrt{g}\,T^\zeta_t\;.
\] 
It is convenient to evaluate 
the r.h.s. of this equation on the brane. Using Eqs.~(\ref{Tvect2}), 
(\ref{Abrane1}) and (\ref{dza3}) one obtains
\[
\d_t {\cal M}=\frac{6\pi^2M^2}{k^3e^2}r_h(\d_t a)^2\;.
\]
Substitution of the relation (\ref{rh}) yields the
following time-dependence of the Schwarzschild radius, 
\be
\label{rha}
r_h=\left(\frac{3\lambda}{2k}\int dt\, (\d_ta)^2\right)^{1/3}\;,
\ee
where we introduced
\be
\label{lambda}
\lambda=\frac{8\pi G_5M^2}{e^2}\;.
\ee
Note that with our assumptions about the hierarchy among the
model parameters discussed in Sec.~\ref{Sec:2}, we have  
$\lambda\sim (M/M_{Pl})^2\ll 1$. 

One proceeds by extracting the functions $\phi$ and $\psi$ defined
in Eq.~(\ref{slow}) from the metric (\ref{adsschw}) and by substituting
them into the effective Friedman equation (\ref{newfried}). One
obtains, 
\[
H^2=\frac{8\pi G_N}{3}(\rho_\Lambda+\rho)+\frac{k^2r_h^4}{a^4}\;,
\]
where we made use of the relation $a=1/(k\zeta_B)$. Finally,
Eq.~(\ref{rha}) gives,
\bseq
\label{Fried}
\be
\label{friedbh1}
H^2=\frac{8\pi G_N}{3}(\rho_\Lambda+\rho)+\delta\;,
\ee
where
\be
\label{addit}
\delta=\frac{k^2}{a^4}
\left(\frac{3\lambda}{2k}\int d\tau\, a \dot a^2\right)^{4/3}\;.
\ee
\eseq
In the last formula we changed the integration variable from the
conformal time  $t$ to the proper time
$\tau$ on the brane using $d\tau=adt$. 

Let us discuss our result. Equations (\ref{Fried})
describe the cosmological evolution in terms of brane variables
alone. The contribution of the bulk is encoded in the term
(\ref{addit}) which is nonlocal in time. 
This feature of our setup makes it different
from other models of dark energy. In particular, the expansion history
of the Universe at the epoch of transition from 
matter to dark energy domination is expected to be distinguishable
from the standard $\Lambda$CDM case. 

The non-locality of Eq.~(\ref{friedbh1}) 
somewhat complicates its exact solution. 
However, the qualitative effect of (\ref{addit})
is easy to work out. Let us first consider the case when $\delta$ is
small and the cosmological evolution is dominated by
matter with equation of state $p=w\rho$, $w<1$. Then one has,
\[
\rho\propto a^{-3(1+w)}\;,
\] 
and
\[
a\propto \tau^{\frac{2}{3(1+w)}}\;.
\]
Substituting this expression 
into Eq.~(\ref{addit}) one notices that
the integral is saturated at the upper limit.
This implies that the choice of the 
integration constant is
irrelevant at the late stages of the cosmological evolution.
One obtains,
\[
\delta=\left(\frac{2\lambda}{3(1-w^2)k}\right)^{4/3}
\frac{k^2}{\tau^{4/3}}
=\left(\frac{\lambda^2 k}{(1-w)^2H}\right)^{2/3}H^2\;.
\]
We see that the $\delta$ term dilutes slowlier than the contribution
of the ordinary 
matter,
\[
\delta\propto a^{-2(1+w)}\;.
\] 
Thus eventually it starts to dominate the cosmological
expansion. To figure out what happens in this case we consider the
opposite limit of small ordinary matter density, 
$\frac{8\pi G_N}{3}\rho\ll~H^2$. The solution to
Eq.~(\ref{friedbh1}) is approximately given by exponential expansion,
$
a\propto \e^{H\tau}
$   
and the Friedman equation takes the form,
\be
\label{friedbh2}
H^2=\frac{8\pi G_N}{3}(\rho_\Lambda+\rho)
+\left(\frac{\lambda^2k}{4H}\right)^{2/3}H^2\;.
\ee
One concludes that as the matter on the brane dilutes the cosmological
expansion 
is attracted to the self--accelerated fixed point. 

\begin{figure}[!htb]
\begin{center}
%\begin{picture}(0,0)%
\includegraphics[width=0.5\textwidth]{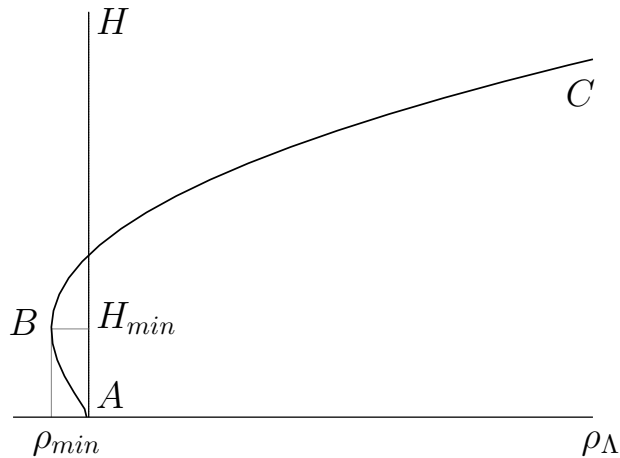}%
%\end{picture}%
\setlength{\unitlength}{3947sp}%
\begingroup\makeatletter\ifx\SetFigFont\undefined%
\gdef\SetFigFont#1#2#3#4#5{%
  \reset@font\fontsize{#1}{#2pt}%
  \fontfamily{#3}\fontseries{#4}\fontshape{#5}%
  \selectfont}%
\fi\endgroup%
%\begin{picture}(5724,6863)(-311,-5787)
\begin{picture}(0,0)(0,0)
{\large
\put(-3250,2500){$H$}
\put(-300,2050){$C$}
\put(-3250,650){$H_{min}$}
\put(-3800,610){$B$}
\put(-200,-100){$\rho_\Lambda$}
\put(-3650,-100){$\rho_{min}$}
\put(-3250,150){$A$}
}
\end{picture}%
\caption{Dependence of the asymptotic Hubble parameter on the
explicit cosmological constant $\rho_\Lambda$. Only the solutions from the
branch $BC$ are stable and can be reached in the course of the
cosmological evolution.}
\label{fig:H}
\end{center}
\end{figure}

The asymptotic Hubble
parameter depends on the explicit cosmological constant 
$\rho_\Lambda$, see
Fig.~\ref{fig:H}. 
For the case $\rho_\Lambda=0$ it takes the value
\[
H_c=\lambda^2k/4\;.
\]
Recalling the assumptions (\ref{assump}) about the parameters of the
model one observes that $H_c$ corresponds to the effective energy
density given by the seesaw formula (\ref{seesaw}).   
The present value of the Hubble parameter is obtained if
$M\sim\text{1 TeV}$. 

We stress that the self-acceleration mechanism described above is not 
equivalent to an implicit introduction of the CC. Indeed, for
$\rho_\Lambda=\rho=0$ equations (\ref{Fried})
admit, apart from the exponentially expanding solution, a static
solution with $H=0$.  However, inspection of Eqs.~(\ref{Fried}) 
shows that this solution is unstable: addition of a
small amount of matter drives it to the self-accelerating
regime. Moreover, one cannot neutralize self-acceleration by
introducing a negative $\rho_\Lambda$. Let us discuss this issue in
more detail. If the matter density $\rho$ is zero and the cosmological
constant is negative but not too large, 
$$
0\geq \rho_\Lambda\geq
\rho_{min}\equiv -\frac{H_c^2}{18\pi G_N}\;,
$$ 
equations (\ref{Fried}) have two branches
of solutions, see Fig.~\ref{fig:H}. However, one can show that the
solutions of the branch $AB$ are unstable: they are destroyed by the
introduction of matter. Thus, only the solutions of the branch $BC$
can be reached in the course of the cosmological evolution. The Hubble
parameter on this branch is bounded from below by
$H_{min}=(2/3)^{3/2}H_c$. If $\rho_\Lambda<\rho_{min}$ there are no solutions
to Eqs.~(\ref{Fried}) unless there is some matter on
the brane, $\rho\neq 0$. We solved Eqs.~(\ref{Fried}) 
numerically in the case of nonrelativistic matter and
$\rho_\Lambda<\rho_{min}$ and found that the Universe eventually
collapses. Thus, the fate of the Universe is restricted to two
possibilities, depending on the value of $\rho_\Lambda$: the Universe
either undergoes a de Sitter phase with the Hubble parameter greater
than $H_{min}$ or collapses.
In other words, if the Universe is not collapsing it is
inevitably self-accelerating.

Let us check various approximations we made in deriving
Eqs.~(\ref{Fried}). Using Eq.~(\ref{rha}) we observe
that the condition (\ref{smallfr}) is satisfied if
\be
\label{smallfr1}
H\ll\sqrt{\lambda}k\;.  
\ee 
For our choice of parameters $\sqrt\lambda
k\sim M$ and the requirement (\ref{smallfr1}) is indeed satisfied for
most stages of the cosmological evolution\footnote{The inequality
(\ref{smallfr1}) holds when matter energy density $\rho$ is smaller
than $(10^9 \text{GeV})^4$.}.  A problem arises with the
quasistationary approximation for the bulk metric. Indeed, from 
Eq.~(\ref{rha}) one concludes that $\d_tr_h/r_h\sim \d_ta/a$ and
therefore the metric evolves at the same rate as the vector fields. 
A
priori, this means that the formulas of this section should be taken
with a grain of salt and are valid only qualitatively. However, we
expect the final result, Eqs.~(\ref{Fried}) to be
correct. This is supported by the explicit solution of the time-dependent
Einstein equations for the case of self-accelerated cosmology in
Sec.~\ref{Sec:4} and the interpretation in terms of AdS/CFT
correspondence in Sec.~\ref{Sec:5}.

%%%%%%%%%%%%%%%%%%%%%%%%%%%%%%%%%%%%%%%%%%%%%%%%%%%%%%%%%%%%%%%%%%%%%%%

\section{Self-accelerated solution}
\label{Sec:4}
In this section we explicitly solve the system of bulk equations for
the metric \eqref{start1} with Ansatz \eqref{start2} 
for the vector fields 
and
show that the model possesses a self-accelerated solution.
It is convenient to write the equations of motion in the
light--cone coordinates
\be
u=k(t-\zeta)~,~~~v=k(t+\zeta)\;.
\ee
A straightforward calculation yields the Einstein equations,
\bseq
\label{einull}
\begin{align}
\label{einull1}
-\frac{\d_v^2r}{r}+\frac{\d_vF\d_vr}{Fr}
&=\frac{\lambda}{r^2}(\d_vA)^2\;,\\
\label{einull2}
-\frac{\d_u^2r}{r}+\frac{\d_uF\d_ur}{Fr}
&=\frac{\lambda}{r^2}(\d_uA)^2\;,\\
\label{einull3}
-\d_u\d_v\ln{F}-\frac{\d_u\d_vr}{r}+\frac{4\d_ur\d_vr}{r^2}
&=\frac{\lambda}{r^2}\d_uA\d_vA\;,\\
\label{einull4}
\frac{\d_u\d_vr}{r}+\frac{2\d_ur\d_vr}{r^2}&=-F\;.
\end{align} 
\eseq
The equation for the vector fields takes the form
\be
\label{venull}
\d_u(r\d_vA)+\d_v(r\d_uA)=0\;.
\ee 
One notices that the system \eqref{einull}, \eqref{venull} is 
invariant under dilatations,
\be
\label{transf}
A(u,v)\mapsto\gamma A(\gamma u,\gamma v)\;,~~~
r(u,v)\mapsto\gamma r(\gamma u,\gamma v)\;,~~~
F(u,v)\mapsto\gamma^2 F(\gamma u,\gamma v)\;,
\ee
where $\gamma$ is an arbitrary constant.
In general, this symmetry is broken by   
embedding of the brane and by the boundary condition (\ref{Abrane}).

An exception from this rule, which is
of primary interest to us, is the case of inflating brane. 
As discussed in Sec.~\ref{Sec:3.1} the metric in the vicinity of the
brane is close to the metric of AdS$_5$, see Eqs.~(\ref{slow}). 
The inflating brane is embedded into the AdS space along the
straight line
\[
\zeta=-\frac{H}{k}t~,~~~t<0\;,
\]
where we used the relation $H\ll k$.
Obviously, this embedding respects the symmetry (\ref{transf}). Due to
the form of the scale factor,
\be
\label{infl}
a=-\frac{1}{Ht}\;,
\ee 
the
boundary condition (\ref{Abrane}) is also invariant under 
the transformations (\ref{transf}).

The general Ansatz invariant under the symmetry (\ref{transf}) has the
form 
\be
\label{ArF}
A=-\frac{1}{\sqrt{\lambda}u}\alpha\left(\frac{v}{u}\right)~,~~~
r=-\frac{1}{u}\beta\left(\frac{v}{u}\right)~,~~~
F=\frac{1}{u^2}f\left(\frac{v}{u}\right)\;.
\ee
Its substitution into Eqs.~(\ref{einull}), (\ref{venull}) reduces
these equations to a system of ordinary differential equations
for the functions $\alpha$, $\beta$, $f$ depending on a single variable
$x=v/u$. One obtains,
\bseq
\label{odiff}
\begin{align}
\label{odiff1}
-\frac{\beta''}{\beta}+\frac{f'\beta'}{f\beta}
&=\left(\frac{\alpha'}{\beta}\right)^2\;,\\
\label{odiff2}
-\frac{\beta''}{\beta}-\frac{2\beta'}{x\beta}+
\frac{f'\beta'}{f\beta}+\frac{f'}{xf}&=
\left(\frac{\alpha'}{\beta}+\frac{\alpha}{x\beta}\right)^2\;,\\
\label{odiff3}
\left(\frac{f'}{f}\right)'+\frac{\beta''}{\beta}+\frac{f'}{xf}
-\frac{2\beta'}{x\beta}-4\left(\frac{\beta'}{\beta}\right)^2
&=-\left(\frac{\alpha'}{\beta}\right)^2-\frac{\alpha\alpha'}{x\beta^2}\;,\\
\label{odiff4}
\frac{\beta''}{\beta}+\frac{4\beta'}{x\beta}
+2\left(\frac{\beta'}{\beta}\right)^2&=\frac{f}{x}\;,\\
\label{odiff5}
\frac{\alpha''}{\alpha}+\frac{\beta'\alpha'}{\beta\alpha}
+\frac{5\alpha'}{2x\alpha}+\frac{\beta'}{2x\beta}&=0\;.
\end{align}
\eseq
The first four equations correspond to the Einstein equations, while the
last one is the equation of motion of the vector fields; only three
equations out of five are independent. One looks for a
solution to these equations in the interval $0<x<1$. The end-points
$x=0$ and $x=1$ of this interval map to 
the horizon
$\zeta=-t$ and
the AdS boundary $\zeta=0$, respectively. 
Note that we again used the inequality $H\ll k$ and identified the
brane with the AdS boundary.\footnote{As a side remark we note that the
system
\eqref{odiff} is valid in the case of $k\sim H$ as well. However, the
boundary conditions~\eqref{odiffbc2} take a more complicated form: in
this case they must be derived directly from the junction conditions
(\ref{junc}) and the difference of the position of
the brane from the AdS boundary must be taken into account.}
We require that the solution
should be regular at the horizon: the functions $\alpha$, $\beta$, $f$
must be continuous at $x=0$.
The value of
the function $\alpha$ at the AdS boundary is fixed by 
Eqs.~(\ref{Abrane1}), (\ref{infl}), (\ref{ArF}):
\be
\label{odiffbc1}
\alpha\to\alpha_0\equiv\frac{\sqrt\lambda k}{H}~,~~~~~x\to 1\;.
\ee
The behavior of the functions $\beta$, $f$ near the boundary is
determined from the requirement that the metric in this region
approximates that of AdS$_5$. Comparing Eqs.~(\ref{AdSmetr}) with
(\ref{ArF}) one obtains,
\be
\label{odiffbc2}
\beta\to\frac{2}{1-x}~,~~~
f\to\frac{4}{(1-x)^2}~,~~~~x\to 1\;.
\ee 
Taking into account Eq.~(\ref{smallfr1}) we consider
the case when the
parameter $\alpha_0$
is large, $\alpha_0\gg 1$. 

The details of the solution are presented in 
appendix~\ref{App:A0}.
Here we list the results. The function $\alpha(x)$ is constant up to
small corrections, $\alpha(x)\approx\alpha_0$. The function
$\beta(x)$ is given implicitly by the following expression,
\be
\label{integr}
-\frac{1}{2\beta_h}
\ln\left[\frac{\beta-\beta_h}{\beta+\beta_h}\right]
-\frac{1}{\beta_h}\left[\arctg\frac{\beta}{\beta_h}
-\frac{\pi}{2}\right]=1-x\;,
\ee
where
\be
\label{betaalpha}
\beta_h=(\alpha_0^2/2)^{1/3}\;.
\ee 
Finally, the function $f(x)$ has the form
\be
\label{frho}
f=\beta^2-\frac{\beta_h^4}{\beta^2}\;
\ee
at $(1-x)\ll 1$ and 
\be
\label{fhor}
f\propto\beta_h^2x^{(2\alpha_0)^{2/3}}
\ee
at $(1-x)\gg 1/\beta_h$. The last expression implies that $f$ tends to
zero at $x=0$, so that this point indeed corresponds to the horizon of
the metric. Note the similarity of the expressions (\ref{integr}),
(\ref{frho}) to the case of the 
AdS-Schwarzschild
metric, Eqs.~(\ref{rzeta}), (\ref{Fr}). 
A careful analysis of the obtained bulk metric shows that it
  indeed describes an AdS$_5$ black hole with growing mass, as
  anticipated by the qualitative arguments of Sec.~\ref{Sec.3.3}.

Once the bulk metric is found one has to substitute it into the
modified Friedman equation (\ref{newfried}): this determines
self-consistently the Hubble parameter of the brane. 
From Eqs.~(\ref{integr}),
(\ref{frho}) 
we find the next-to-leading order asymptotics of the functions
$\beta$ and $f$ at $x\to 1$, 
\begin{align*}
&\beta=\frac{2}{1-x}\left(1+\frac{\beta_h^4}{80}(1-x)^4\right)\;,\\
&f=\frac{4}{(1-x)^2}\left(1-\frac{3\beta_h^4}{80}(1-x)^4\right)\;.
\end{align*}
Substitution of these expressions into Eqs.~(\ref{ArF})
yields that the functions $r$ and $F$ have the form \eqref{slow} in
the vicinity of the brane with
\[
\psi=\frac{\beta_h^4}{5}\left(\frac{\zeta}{t}\right)^4\;,~~~~~~~~
\phi=-\frac{3\beta_h^4}{10}\left(\frac{\zeta}{t}\right)^4\;.
\]
Then from Eq.~(\ref{newfried}) we obtain the modified Friedman
equation (\ref{friedbh2}).  

The following comment is in order. So far
in this section we have assumed that the brane is completely empty, so
that its expansion is purely inflationary. However, the
above analysis is also applicable
if there is
matter on the brane provided that the matter density is small,
$G_N\rho\ll H^2$.

%%%%%%%%%%%%%%%%%%%%%%%%%%%%%%%%%%%%%%%%%%%%%%%%%%%%%%%%%%%%%%%%%%%%%%%%%

\section{AdS/CFT interpretation}
\label{Sec:5}

To understand better the mechanism of self--acceleration
it is useful to consider it from the point of view of the
AdS/CFT  
interpretation of the RS model \cite{ArkaniHamed:2000ds}. 
In this
framework the 5-dimensional setup considered above is dual to the
4-dimensional theory described by the action (cf. Eq.~(\ref{Sbrane}))
\[
S_{4d}=\int d^4x\sqrt{-g}
\left(-\frac{R}{16\pi G_N}+{\cal L}_{SM}+
{\cal L}_{HS}[A_\mu^a,\ldots]+{\cal L}_{CFT}[A^a_\mu,\ldots]\right)\;,
\]  
where ${\cal L}_{CFT}$ represents the Lagrangian of a
strongly interacting conformal
field theory coupled to the vector fields. 
The properties of this CFT are determined by the
structure of the bulk sector of the 5d theory. 
The expression for the
energy density of the CFT plasma with temperature $T$ has the
following form\footnote{The expression (\ref{rhoCFT}) differs by a
  factor 2 from that of Ref.~\cite{Gubser:1999vj}. This is related to
  the fact that in contrast to \cite{Gubser:1999vj} where the brane is
  considered as the ``end-of-the-universe'', we assume the brane to be
  embedded into an 
  infinite space-time consisting of two copies of AdS related by the
  $Z_2$ symmetry. This leads to the doubling of the
CFT degrees of freedom as compared to \cite{Gubser:1999vj}.} 
\cite{Gubser:1999vj},
\be
\label{rhoCFT}
\rho_{CFT}=\frac{3\pi^3}{8G_5k^3}T^4\;.
\ee
The presence of the $[U(1)]^3$ gauge symmetry in the five-dimensional
Lagrangian
implies
that the CFT is coupled via $[U(1)]^3$ 
gauge interactions to the vector
fields $A^a_\mu$. The latter are identified with the values of the
vector fields on the brane in the 5d setup.

In the 4d language the condition (\ref{ABC}) leads to the following
cosmological Ansatz for the 
vector fields:
$A^a_0=0$, $A_i^a=Ma(t)\delta_i^a$. This configuration corresponds to
three orthogonal homogeneous ``electric'' fields\footnote{To avoid
  confusion we stress
  that these are the fields of the hidden sector and have nothing to
  do with the electromagnetic field of the Standard Model.}; 
in the local inertial
frame they have the form,
\be
\label{electric}
E^a_i=MH\delta^a_i\;.
\ee
These electric fields give rise to electric currents ${\bf j}^a$ in the CFT
plasma. 
In the linear response approximation the currents 
read 
\be
\label{current}
{\bf j}^a=c_1\dot {\bf E}^a+c_2 T{\bf E}^a\;,
\ee 
where $c_1$, $c_2$ are dimensionless coefficients. 
The expression \eqref{current} is the most general one allowed by dimensional
considerations. 
The coefficients $c_1$, $c_2$
can be extracted from the study of the vector
fields in AdS, see appendix~\ref{App:A}. The result is 
\bseq
\label{c}
\begin{align}
\label{c1}
&c_1=\begin{cases}
\frac{2}{e^2k}\ln\left[\frac{k}{\Omega}\right]~,~~~~~\Omega\gg T\\
\frac{2}{e^2k}\ln\left[\frac{k}{T}\right]~,~~~~~\Omega\ll T
\end{cases}\;,\\
\label{c2}
&c_2=\frac{2\pi}{e^2k}\;,
\end{align}
\eseq
where $\Omega$ is the characteristic frequency of the electric field.
For the case of cosmology we have $\Omega\sim H\ll T$ and the first
term on r.h.s. of Eq.~(\ref{current}) can be omitted. 

The currents heat the plasma with the
energy release per unit time per unit volume given by 
\be
\label{pilup}
W=\sum_a {\bf j}^a{\bf E}^a\;.
\ee
Thus, 
the energy balance equation for the CFT plasma has the form
\be
\label{ebalance}
\dot\rho_{CFT}+4H\rho_{CFT}=W\;.
\ee 
Substituting expressions (\ref{rhoCFT}), (\ref{pilup}),
(\ref{electric}) into Eq. \eqref{ebalance}, 
and solving the resulting differential equation for 
the temperature of the plasma we obtain
\[
T=\frac{1}{a}\left(\frac{3\lambda k^2}{2\pi^3}
\int d\tau\,a\dot a^2\right)^{1/3}\;,
\]
where $\lambda$ is defined in (\ref{lambda}). 
One observes that the contribution 
\[
\delta=\frac{8\pi G_N}{3}\rho_{CFT}
\]
of the CFT plasma into the 4-dimensional Friedman equation
coincides with the expression~(\ref{addit}). 

The analysis of this section suggests a transparent interpretation of
the self--acceleration mechanism in our model. It is based on the
inflow of energy into the conformal matter which compensates for
cooling of the plasma due to the cosmological expansion. Ultimately,
this leads to stabilization of the plasma temperature.  Similar ideas
were expressed previously in Ref.~\cite{Apostolopoulos:2005at}; our
model provides explicit realization of these ideas in the framework of
field theory.  Note that the CFT picture also indicates that
Eqs.~(\ref{Fried}) 
are more robust than could be inferred from the
5-dimensional analysis of Sec.~\ref{Sec:3}. Indeed, as discussed in
Sec.~\ref{Sec:3}, the 5d analysis is not completely reliable as it
does not self-consistently take into account the time evolution of the
bulk metric.  On the other hand, the formulas of the present section
are expected to be valid once the CFT plasma is  
in thermal equilibrium. This requirement is satisfied if the Hubble
time $1/H$ is much larger than the thermalization time of the
plasma. By dimensionality, the latter is of order
$1/T$. As $1/T\ll 1/H$, we see that the condition of thermal
equilibrium is always satisfied.

In our model the inflow of energy into the plasma is produced by the
work done by electric fields. The key point here is to
find a way to prevent the electric fields from rapid decay which
is usually caused by the cosmological expansion.
The condition (\ref{ABC}) is crucial for this purpose.

This consideration reveals the following property: the
hidden sector responsible for the condition (\ref{ABC}) has to violate
null energy condition (NEC). Indeed, the equation of the energy
balance for the hidden sector implies
\[
\dot\rho_{HS}+3H(\rho_{HS}+p_{HS})=-W\;,
\] 
where $W$ is given by Eq.~(\ref{pilup}).
As $W>0$ and $\dot\rho_{HS}\to 0$ at $\tau\to+\infty$ (cf. condition (2)
of Sec.~\ref{Sec:2}) the NEC is violated:
\be
\label{NECHS}
\rho_{HS}+p_{HS}<0\;.
\ee
Generically, violation of NEC leads to instabilities
\cite{Dubovsky:2005xd}, so the condition (\ref{NECHS}) is rather
worrisome. We discuss this issue in more detail in the next section.   

Let us point out possible phenomenological generalizations
of the self-accele\-ration mechanism proposed in this paper. One can
consider a class of cosmic fluids interacting with electric fields
(\ref{electric}) with a general dependence of the conductivity on the
density of the fluid,
\[
{\bf j}^a=c(\rho_f){\bf E}^a\;.
\]
The case considered in this paper corresponds to $c\propto \rho^{1/4}_f$.
The equation of the energy balance of the fluid takes the form 
\be
\label{balancegen}
\dot{\rho}_f+3H\big(\rho_f+p_f\big)=3M^2H^2c(\rho_f) \;.
\ee
Again, one assumes that the constant electric fields (\ref{electric})
are produced by some NEC-violating hidden sector, whose energy density
$\rho_{HS}$ is zero (or tends to zero during the cosmological expansion).
Such a setup will lead to self-accelerated cosmology
if the system consisting of Eq.~(\ref{balancegen}) and the Friedman
equation has an attractor fixed point with $\rho_f\neq 0$. We point
out, however, that we are not aware of a microscopic theory of a
fluid with $c(\rho_f)$ different from $c\propto \rho_f^{1/4}$.

%%%%%%%%%%%%%%%%%%%%%%%%%%%%%%%%%%%%%%%%%%%%%%%%%%%%%%%%%%%%%%%%%%%%%%%%%

\section{Discussion}
\label{Sec:7}

In this paper we proposed a model realizing the self-accelerated
cosmology.  The model is formulated in the context of a brane-world
scenario with warped extra dimensions.  We have shown that the regime
of accelerated cosmological expansion is an attractor in the case when
there is a sufficient energy exchange between the expanding brane and
the bulk.  This exchange is realized by means of bulk vector fields
which have a specific interaction with the brane. The key requirement
for the existence of the self-accelerated solution is that this
interaction produces the boundary condition (\ref{ABC}), which relates
the vector fields to the scale factor. The condition (\ref{ABC}) is
analogous to the conditions appearing in effective $\sigma$-models of
particle physics.

The condition (\ref{ABC}) is far from being trivial: it breaks
explicitly the (Abelian) gauge symmetries of the bulk action;
moreover, it leads to spontaneous breaking of the Lorentz symmetry. Per
se, these properties do not imply any inconsistencies. In particular,
there exist a number of self-consistent models
\cite{ArkaniHamed:2003uy,Dubovsky:2004sg,Libanov:2005vu} involving
spontaneous Lorentz symmetry breaking. Still, the stability
requirements impose tight constraints on the structure on these
models.  In our case construction of a suitable brane Lagrangian
${\cal L}_{HS}$ is complicated by the fact that it has to produce an
energy--momentum tensor violating NEC.

Let us demonstrate that, under quite generic
assumptions, the latter requirement is unavoidable in
brane-world models with self-accelerated cosmology. Indeed, 
consider a brane which performs de Sitter expansion in a
higher-dimensional bulk. The energy conservation equation for the
brane has the form,
\[
\dot\rho_b+3H(\rho_b+p_b)=-W\;,
\]
where $\rho_b$, $p_b$ are the density and pressure of the matter on
the brane and $W$ is the energy flux radiated by the brane into the
bulk. Let us assume that the brane energy--momentum tensor respects the
symmetries of de Sitter space-time. This implies
$\dot\rho_b=0$. Another natural assumption is that $W\geq 0$: the brane
{\em radiates} rather than absorbs energy from the bulk. Thus we
obtain
\be
\label{NECbrane}
\rho_b+p_b\leq 0\;.  
\ee 
The equality can be attained only if $W=0$,
i.e. the brane--bulk energy exchange is absent. A self-accelerated
cosmology of this type is not excluded: an example is provided by the
DGP model \cite{Dvali:2000hr}. However, generically one expects $W>0$
(this is the case in our model). Then, the inequality in
(\ref{NECbrane}) is strong and the brane energy--momentum tensor
violates NEC.

It was shown in Ref.~\cite{Dubovsky:2005xd} that under broad
assumptions violation of NEC leads to exponential instabilities. At
first sight, this statement precludes any hope to construct a viable
brane Lagrangian fitting into the self-acceleration mechanism proposed
in this paper. However, important assumptions of
\cite{Dubovsky:2005xd} are that the NEC violating energy--momentum
tensor is associated with {\em dynamical fields} and that one can
neglect higher derivative terms in the action of these fields. This
suggests two possible ways to avoid instabilities.
First, one can envisage that higher derivative terms in the action for
unstable modes cut off the instabilities at short wavelengths
\cite{Creminelli:2006xe} and make
them slow enough to satisfy phenomenological 
constraints. Second,
some or all of the fields localized on the brane may be constrained to
vanish by additional requirements which should be attributed to
UV-completion of the theory.  In appendix \ref{App:A00} we present a
model brane Lagrangian exploiting the latter possibility. For this
Lagrangian we have been able to establish, in a number of
phenomenologically relevant regimes, the absence of instabilities
predicted by \cite{Dubovsky:2005xd}. The details of this stability
analysis will be published in a separate article \cite{Perturbations}.

The model considered in this paper requires UV-completion above the
energy scale $M\sim 1$~TeV. The latter scale turns out to be the same
as the natural scale for the new physics stabilizing the Standard
Model Higgs mass; so, it is tempting to speculate that the two scales
could have common physical origin. In this case our seesaw formula
(\ref{seesaw}) for the effective energy density corresponding to the
late-time cosmological expansion would look appealing in light of the
cosmic coincidence problem --- the approximate equality of the
densities of dark energy and dark matter at the present epoch. Indeed,
as shown in \cite{ArkaniHamed:2000tc}, the cosmic coincidence is a
natural consequence of Eq.~(\ref{seesaw}) in the case when the dark
matter consists of massive particles whose masses and interactions are
set by the scale $M$. Particles with the latter properties are generic
in the UV extensions of the Standard Model.

It would be interesting to embed our setup in some higher
dimensional (super)gravity theory. The following considerations
suggest that this may be possible. The $[U(1)]^3$ gauge symmetry of
the bulk action may be interpreted as the isometry of the internal
space in the compactification of 8-dimensional gravity on a
3-dimensional torus. The bulk vector fields then naturally appear from
the components of the 8-dimensional metric.  The breaking of the
$U(1)$ symmetries on the brane is also natural and corresponds to
breaking of the translational symmetry of the internal torus by the
position of the brane in it. In this context the condition (\ref{ABC})
should appear as the consequence of nontrivial winding of the 3-brane
around the internal torus. We leave the development of this idea for
future studies.
  
We have considered interpretation of our model in terms of AdS/CFT
correspondence. The CFT interpretation of the self-acceleration
mechanism proposed in this paper allows to formulate its
phenomenological generalizations. The latter involve consideration of
a class of cosmic fluids interacting with ``electric'' fields.  It
would be interesting to work out predictions of this class of models
and confront them with observations.

As a by-product of our study we found an explicit solution of the
Einstein equations describing a black hole with time-dependent mass in
AdS$_5$.  We used this solution to embed a brane into it. However,
this solution can be considered on its own without introduction of the
brane. In this case the solution is dual by AdS/CFT correspondence to a
time-dependent configuration of CFT in {\it flat} 4d space-time. One
finds that it provides a self-consistent description of the response
of CFT plasma to a triplet of external time-dependent electric
fields\footnote{In this context $t$ is the physical time and the value
of the electric fields should be taken on the AdS boundary.}
$$
E^a(t)\propto 1/t^2\;.
$$
It would be interesting to understand whether this solution can be
useful for the description of collective phenomena in strong coupling
regime of gauge theories. We leave this question for future.

{\bf Acknowledgments.}  We are indebted to F.~Bezrukov, S.~Dubovsky,
M.~Libanov, P.~Koroteev, V.~Rubakov, A.~Boyarsky and O.~Ruchayskiy for
useful discussions. We thank Service de Physique Th\'eorique de
l'Universit\'e Libre de Bruxelles for hospitality.  This work was
supported in part by the grants of the President of the Russian
Federation NS-7293.2006.2 (government contract 02.445.11.7370) and
MK-1957.2008.2 (DG), by the RFBR grant 08-02-00473-a (DG), by the
Russian Science Support Foundation (DG) and by the EU 6th Framework
Marie Curie Research and Training network "UniverseNet"
(MRTN-CT-2006-035863) (SS).

%%%%%%%%%%%%%%%%%%%%%%%%%%%%%%%%%%%%%%%%%%%%%%%%%%%%%%%%%%%%%%%%%%%%%%%%%%%%%%%%%

\appendix

%%%%%%%%%%%%%%%%%%%%%%%%%%%%%%%%%%%%%%%%%%%%%%%%%%%%%%%%%%%%%%%%%%%%%%%%%%%%%%%%%%

\section{An example of the brane sector Lagrangian}
\label{App:A00}
As a possible choice for the Lagrangian of the brane sector of our
model we
propose
\be
\label{LHS}
{\cal L}_{HS}=\chi_1(\bar g^{\mu\nu}\d_\mu\phi\d_\nu\phi-1)
+\chi^{ab}(\bar g^{\mu\nu}\bar A^a_\mu \bar A^b_\nu+M^2\delta^{ab})\;,
\ee
where 
\[
\bar A^a_\mu=A^a_\mu-\bar g^{\nu\lambda}A^a_\nu\d_\lambda\phi
\d_\mu\phi
\]
and $\chi_1$, $\chi^{ab}$ are Lagrange multipliers. Note that the
Lagrange multiplier $\chi_1$ enforces the gradient of the field $\phi$
to be a unit time-like vector. This implies that the field $\phi$ is
non-dynamical; it plays the role of ``cosmic
time''. 

Let us demonstrate that the hidden sector (\ref{LHS}) satisfies the 
conditions (1), (2) of Sec.~\ref{Sec:2}. In the cosmological context
one makes an Ansatz
\[
\dot\phi=1~,~~~\d_i\phi=0\;.
\]
Combining with Eqs.~(\ref{start2}), (\ref{brmetric}) we see that in
the cosmological setting $\bar A^a_\mu=A^a_\mu$. Then, the Lagrange
multiplier $\chi^{ab}$ enforces the condition (1). From (\ref{LHS})
one obtains the energy--momentum tensor of the hidden sector,
\[
T_{\mu\nu}^{(HS)}=2\chi_1\d_\mu\phi\d_\nu\phi+
2\chi^{ab} \bar A^a_\mu \bar A^b_\nu\;.
\]
For the cosmological Ansatz
\[
\rho_{HS}=2\chi_1\;.
\]
The evolution of $\chi_1$ is determined from the equation obtained by
varying (\ref{LHS}) with respect to $\phi$,
\[
\nabla^\mu(\chi_1\d_\mu\phi-
\chi^{ab}\bar A^a_\mu A^{b\,\nu}\d_\nu\phi)=0\;.
\]
For the cosmological setup this equation simplifies and yields
\[
\chi_1\propto\frac{1}{a^3}\;.
\] 
Thus the condition (2) is also satisfied.

As discussed in the main text the energy--momentum
tensor of the hidden sector violates NEC on the self-accelerated
solution, so one has to worry about possible instability of this background
due to rapidly growing short wavelength modes
\cite{Dubovsky:2005xd}. We have
carried out an analysis of stability of the self-accelerated
background in the model with the hidden sector Lagrangian (\ref{LHS});
this analysis will be presented in
Ref.~\cite{Perturbations}. The preliminary results of this analysis
are that the
instabilities are absent in the range of wavelengths/time-scales 
\be
\label{range}
(HM_{Pl})^{-1/2}\lesssim \Delta l,~\Delta\tau\lesssim H^{-1}\;. 
\ee 
We also found that in this range of distances/time-intervals the
linearized gravitational field of external sources is described by the
standard formulas of the 
four-dimensional Einstein's gravity. 

The range (\ref{range}) of distances/time-scales, where we were able to
establish the absence of instabilities, is limited both on
short- and long-scale sides. This is due to technical difficulties of
the analysis and does not imply that instabilities appear outside this
range. On the other hand, for the moment, we are not able to
completely eliminate this possibility.

\section{Self-accelerated solution: technicalities}
\label{App:A0}

Here we construct the solution of the system of ordinary differential
equations (\ref{odiff}) with boundary conditions (\ref{odiffbc1}),
(\ref{odiffbc2}). 
First, we consider the system (\ref{odiff}) in the
interval $x\in [1-\epsilon, 1]$, where $\epsilon$ is some small
number, $\epsilon\ll 1$, to be specified later. In this region one can
remove the explicit dependence on $x$ in the equations by setting 
$x=1$.  
Multiplying Eq.~(\ref{odiff1}) by a
factor of 2 and subtracting it from Eq.~(\ref{odiff3}) one obtains
\[
\left(\frac{f'}{f}\right)'-\frac{f'\beta'}{f\beta}
+\frac{2\beta''}{\beta}-4\left(\frac{\beta'}{\beta}\right)^2=
-2\left(\frac{\alpha'}{\beta}\right)^2-\frac{\alpha\alpha'}{\beta^2}
-\frac{f'}{f}+\frac{2\beta'}{\beta}\;.
\] 
Divided by $\beta$, this equation can be integrated as 
\be
\label{interm1}
\frac{f'}{f\beta}+2\frac{\beta'}{\beta^2}=C
+\int dx \left(-2\frac{(\alpha')^2}{\beta^3}-\frac{\alpha\alpha'}{\beta^3}
-\frac{f'}{f\beta}+\frac{2\beta'}{\beta^2}\right)\;,
\ee
where $C$ is the integration constant to be determined from the
boundary conditions. We will shortly see that the
terms in the integrand in Eq.~(\ref{interm1}) are of the order one or
smaller. As $x$ changes in the narrow interval of the length
$\epsilon\ll 1$
the whole integral is small and can be neglected. Then, using the
asymptotics (\ref{odiffbc2}) we obtain,
\be
\label{firstint}
\frac{f'}{f\beta}+2\frac{\beta'}{\beta^2}=2\;.
\ee 
Consider now 
Eq.~(\ref{odiff5}). It can be integrated with the result
\be
\label{secondint}
\beta\left(\alpha'+\frac{\alpha}{2}\right)=
\frac{\beta^4_h}{\alpha_0}-\int dx\,2\beta\alpha'=
\frac{\beta_h^4}{\alpha_0}+O(\beta\alpha'\epsilon)\;,
\ee
where $\beta_h$ is an integration constant and we explicitly wrote
down the corrections to keep them under control. 
Substituting the expressions (\ref{firstint}), (\ref{secondint})
into the difference of equations (\ref{odiff2}) and (\ref{odiff1}) we
obtain
\be
\label{rhoeq}
-\frac{2\beta'}{\beta}+\beta=\frac{\alpha \beta_h^4}{\alpha_0\beta^3} 
+O\left(\frac{\alpha\alpha'\epsilon}{\beta^2}\right)=
\frac{\beta_h^4}{\beta^3}
+O\left(\frac{\alpha'\beta_h^4\epsilon}{\alpha_0\beta^3}\right)
+O\left(\frac{\alpha\alpha'\epsilon}{\beta^2}\right)\;,
\ee
where in the last expression we singled out the value of the function
$\alpha$ on the boundary and accounted for the corresponding
corrections. Neglecting the corrections in Eq.~(\ref{rhoeq}) we can
integrate it with the result (\ref{integr}).
Then, from Eq.~(\ref{odiff4}) we obtain the formula (\ref{frho}) 
up to relative 
corrections of order $O(1/\beta)$.
Now, one can check that the
equations (\ref{firstint}), (\ref{odiff1}) are satisfied, the latter
up to corrections $O((\alpha'/\beta)^2)$.
Let us consider the asymptotic behavior of the metric functions
when $\beta\approx\beta_h$. From Eqs.~(\ref{integr}), (\ref{frho}) we
obtain,
\be
\label{asymp}
\beta-\beta_h\propto\beta_h\e^{-2\beta_h(1-x)}~,~~~~
f\propto\beta_h^2\e^{-2\beta_h(1-x)}\;.
\ee
We see that the functions $\beta$ and $f$ exponentially approach
their asymptotics at $1-x\gg 1/\beta_h$. We will shortly see that
$\beta_h\gg 1$. Consequently, the parameter $\epsilon$ introduced
above can be chosen to satisfy $1/\beta_h\ll\epsilon\ll 1$. Finally,
from Eq.~(\ref{secondint}) we obtain at 
$1/\beta_h\ll 1-x\ll \epsilon$ 
\be
\label{alphaprime}
\alpha'\approx \frac{\beta_h^3}{\alpha_0}-\frac{\alpha_0}{2}\;,~~~~~
\alpha\approx\alpha_0\;,
\ee 
where we neglected corrections of order $O(\alpha'\epsilon)$.

We now solve the system (\ref{odiff}) for the values of $x$ which are 
well
separated from the boundary, 
$1-x\gg 1/\beta_h$. In this
region $f\ll 1$. One sets $f=0$ in 
Eq.~(\ref{odiff4}) and integrates with the result 
\be
\label{rhohor}
\beta=\left(C_1+\frac{C_2}{x^3}\right)^{1/3}\;.
\ee  
One chooses the solution which is regular at $x=0$. 
Thus, we obtain that $\beta$ is
simply constant, $\beta=\beta_h$. 
Inserting this result into Eq.~(\ref{odiff5}) we obtain
\[
\alpha=D_1+\frac{D_2}{x^{3/2}}\;.
\]
Again, we should take the constant solution,
$\alpha=\alpha_0$. Substituting $\alpha'=0$ into 
Eq.~(\ref{alphaprime}) we obtain the relation (\ref{betaalpha}).
Thus, $\beta_h$ is indeed large, $\beta_h\gg 1$.
Equation (\ref{odiff1}) is now trivially satisfied while 
Eqs.~(\ref{odiff2}), (\ref{odiff3}) reduce to
\[
\frac{f'}{f}=\frac{\alpha_h^2}{x\beta_h^2}\;.
\]
After integration this gives Eq.~(\ref{fhor}); the latter matches with 
Eq.~(\ref{asymp})
provided the relation 
(\ref{betaalpha}) is satisfied.

Now, one can go over the derivation and check that the corrections which
we neglected in various expressions are small. For example, consider
the corrections in Eq.~(\ref{rhoeq}). From the relation
(\ref{betaalpha}) it follows that $\alpha'\lesssim\alpha_0$; taking
into account that $\beta >\beta_h$ one obtains
$\alpha\alpha'\epsilon/\beta^2 \ll\beta$. 

We have checked the analysis described in this appendix by
numerical solution of the system (\ref{odiff}) with boundary
conditions (\ref{odiffbc1}), (\ref{odiffbc2}).

%%%%%%%%%%%%%%%%%%%%%%%%%%%%%%%%%%%%%%%%%%%%%%%%%%%%%%%%%%%%%%%%%%%%%

\section{Conductivity of CFT matter}
\label{App:A}

In this appendix we calculate the conductivity of the CFT plasma which
is dual to the bulk sector of our model. A similar calculation for the
${\cal N}=4$ supersymmetric Yang--Mills plasma was performed in
Ref.~\cite{CaronHuot:2006te}.  First, let us relate the temperature of
the plasma to the properties of the bulk geometry. By the standard
rules of the AdS/CFT duality (see, e.g., Ref.~\cite{Aharony:1999ti})
the finite temperature on the CFT side corresponds to the presence of
a black hole in the bulk. The temperature of the plasma is then equal
to the Hawking temperature of the black hole~\cite{Gubser:1999vj}, 
\[
T=\frac{kr_h}{\pi r}\;.
\]

Second, we use the relation (\ref{pilup}) to derive the formula for
the CFT current. Equating $W$ to the energy 
injected by the brane into the bulk,
$W=2T^\zeta_t$ (where a factor of 2 is due to $Z_2$-symmetry)  
and using Eq.~(\ref{Tvect2}) we obtain,
\[
{\bf j} {\bf E}=-\frac{2M^2}{e^2Fr^2}\d_\zeta A\d_tA\;.
\]
Here we canceled the factor 3 appearing both in Eqs.~(\ref{Tvect2}) and
(\ref{pilup}) due to the presence of 3 vector fields. Using the
relation 
\[
E=M\d_tA/a^2
\]
between the physical electric field and the vector potential we obtain
\be
\label{j}
j=-\frac{2M}{Fe^2}\d_\zeta A\;.
\ee
This formula expresses the CFT current in terms of the quantities
describing the response of the bulk system to the boundary
perturbations. 

Let us consider periodic electric field,
$E=E_0\e^{-i\Omega\tau}$. This corresponds to periodic perturbation of
the vector fields on the AdS boundary.
Note that the physical frequency $\Omega$ is
related to the frequency $\omega$ in conformal time $t$ by $\Omega=\omega/a$.
The response of the bulk to such perturbation was calculated 
in Sec.~\ref{Sec:3}. Let us
first consider the case of small frequency, $\Omega\ll T$. It is easy
to see that this condition is equivalent to Eq.~(\ref{smallfr}), thus
we can use the formula (\ref{dza2}). Substitution of the latter
formula into Eq.~(\ref{j}) yields
\[
j=\left\{-\frac{i\Omega}{2e^2k}\ln\left[\left(\frac{k}{\pi T}\right)^4
-1\right]+\frac{2\pi}{e^2k}\right\}\frac{r^2}{F}E\;.
\]
Taking $T\ll k$ and using $r^2/F\approx 1$ on the brane one
obtains the expressions (\ref{c1}) (lower case) and (\ref{c2}). In the
opposite limit $\Omega\gg T$ the response of the vector field to the
boundary perturbation is not affected by the presence of the black
hole. In this case we can use Eq.~(\ref{solution}) for the propagation
of the vector field in the AdS metric. Performing the same steps as
above we obtain
\[
j=\frac{\pi\Omega}{e^2k}H_0^{(1)}\left(\frac{\Omega}{k}\right)E\;.
\] 
This reduces to Eq.~(\ref{c1}) (upper case) in the regime 
$\Omega\ll k$.

%%%%%%%%%%%%%%%%%%%%%%%%%%%%%%%%%%%%%%%%%%%%%%%%%%%%%%%%%%%%%%%%%%%%
%%%%%%%%%%%%%%%%%%%%%%%%%%%%%%%%%%%%%%%%%%%%%%%%%%%%%%%%%%%%%%%%%%%

\end{document}